\documentclass[aps,amssymb,prb,twocolumn]{revtex4}
\usepackage{graphicx}
\usepackage{epsfig,amsmath}

\begin{document}

\title{Photon heat transport in low-dimensional nanostructures }

\author{Teemu Ojanen}
\email[Correspondence to ]{teemuo@boojum.hut.fi}
\author{Tero T.~Heikkil\"a}
\affiliation{ Low Temperature Laboratory, Helsinki University of
Technology, P.~O.~Box 2200, FIN-02015 HUT, Finland }

\date{\today}
\begin{abstract}
At low temperatures when the phonon modes are effectively frozen,
photon transport is the dominating mechanism of thermal relaxation
in metallic systems. Starting from a microscopic many-body
Hamiltonian, we develop a nonequilibrium Green's function method to
study energy transport by photons in nanostructures. A formally
exact expression for the energy current between a metallic island
and a one-dimensional electromagnetic field is obtained. From this
expression we derive the quantized thermal conductance as well as
show how the results can be generalized to nonequilibrium
situations. Generally, the frequency-dependent current noise of the
island electrons determines the energy transfer rate.

\end{abstract}
\pacs{PACS numbers: } \bigskip

\maketitle

General physical and information-theoretic arguments imply that
there is a fundamental limit $G_Q=\pi^2k_\mathrm{B}^2T/3h$ to the
thermal conductance of a single channel \cite{pendry},
independent of the nature of the conduction mechanism. 
Particularly, $G_Q$ should be independent of the dispersion relation
and quantum statistics of carrier particles.\cite{rego} The
few-channel heat conductance is particularly relevant in
low-dimensional nanostructures where the channel number is naturally
low. The quantized thermal conductance has been experimentally
verified for electrons, phonons and recently in photon transport
between metallic islands.\cite{chiatti,schwab,meschke} If electron
transport is restricted and the system is at very low temperature so
that the phonon modes appear frozen, the dominant thermal relaxation
process is photon transport.\cite{schmidt,meschke}

In this Letter we study energy transport by photons between a
metallic island and a one-dimensional electromagnetic field
supported by a transmission line. The latter mimics the effect of
the external leads and connectors on the island. Our aim is to give
the photon transport a microscopic basis as well as to study
nonequilibrium processes. Applying Green's function methods to a
microscopic model, we obtain a formally exact expression for the
energy current. We show that frequency-dependent current noise
determines the characteristics of the transport process, thus
providing a close connection between the electrons and the photons.
We derive an expression for the heat flow between the field and the
island and verify that the maximum value of thermal conductance in
the system is $G_Q$. This provides a microscopic description of the
recent experiment on electron-photon coupling.\cite{meschke} We
consider also a many-channel case where the electron system is
connected to several transmission lines. The energy current formula
allows us to study situations where the island is driven to a
nonequilibrium state and examine how electron shot noise alters the
energy transport. We show that, due to shot noise, part of Joule
heat flows to the photons. Our results have relevance in determining
the electron temperature of driven mesoscopic systems
\cite{giazotto}, but they are also important in photon-based
solid-state applications, such as cavity QED and its quantum
information realizations.\cite{wallraff, blais}

\begin{figure}[h]
\centering
\includegraphics[width=0.73\columnwidth]{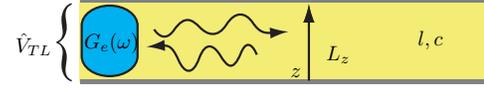}
\caption{(Color online) Metallic island (blue) coupled to the
electromagnetic field of a transmission line (lines around the yellow region). 
The voltage between the strips is
$\hat{V}_{\mathrm{TL}}$.} \label{setuppi}
\end{figure}
The studied model consists of a small metallic island coupled to a
parallel strip transmission line, see Fig. \ref{setuppi}. The
transmission line acts as a waveguide supporting a one-dimensional
electromagnetic field. In contrast to three-dimensional waveguides,
the parallel strip line field has only one allowed field
polarization. Therefore it corresponds to a single transport
channel. There is no direct electrical connection between the
electrons on the island and those in the transmission line strips,
only the field couples to the electrons. The total Hamiltonian of
the system is $H=H_{\mathrm{e}}+H_{\gamma}+H_{\mathrm{e-\gamma}}$,
where
\begin{align}
H_{\mathrm{e}}=&\int\hat{\Psi}^{\dagger}(r)\left(\frac{\hat{p}^2}{2m}+U(r)\right)\hat{\Psi}(r)dr+\nonumber\\
&+\frac{1}{2}\int\hat{\Psi}^{\dagger}(r)\hat{\Psi}^{\dagger}(r')V(r,r')\hat{\Psi}(r')\hat{\Psi}(r)drdr',\\
H_{\gamma}=&\sum_j\hbar\omega_j(\hat{a}_j^{\dagger}\hat{a}_j+\frac{1}{2})\label{fotoni}\\
H_{\mathrm{e-\gamma}}&=g\int\hat{\Psi}^{\dagger}(r)z\hat{\Psi}(r)dr\hat{V}_{\mathrm{TL}}.
\label{coupl}
\end{align}
In the following, 
we do not have to specify the terms in the electron Hamiltonian
$H_e$ in more detail. The transmission line is characterized by its
length $L$, distance between the parallel strips $L_z$, and
inductance $l$ and capacitance $c$ per unit length. Operator
$\hat{V}_{\mathrm{TL}}=\sum_j T_j (\hat{a}_j+\hat{a}_j^\dagger)$ is
the voltage operator at the end of the line. The integration in
$H_{\mathrm{e-\gamma}}$ is restricted to the region between the
parallel strips in the case the electron system extends beyond that.
The island is assumed to be much smaller than the photon wavelength
at the relevant frequencies, so the position dependence of the
voltage operator in the interaction term can be neglected. The field
operators $\hat{\Psi}(r), \hat{\Psi}^{\dagger}(r)$
 and creation and annihilation operators
$\hat{a}_j, \hat{a}_j^{\dagger}$ satisfy canonical fermion and boson
commutation relations. Constants $\omega_j=j\pi v/L$ ($j$ is a
positive integer), $T_j=\sqrt{\hbar\omega_j/Lc}$ and $g=e/L_z$ can
be found by quantizing the line field.\cite{blais} The wave velocity
$v$ in the transmission line is given by $v=1/\sqrt{lc}$.

The electron Hamiltonian $H_{\mathrm{e}}$ does not commute with the
total Hamiltonian $H$, thus there is an energy flow between the
island and the field. This energy flow into the island is
characterized by a current $J_Q$ defined as
\begin{align}\label{cur}
J_Q\equiv\langle\dot{H}_{\mathrm{e}}\rangle= -\frac{g}{m}
\left\langle\int\hat{\Psi}^{\dagger}(r)\hat{p}_z\hat{\Psi}(r)dr\hat{V}_{\mathrm{TL}}\right\rangle.
\end{align}
The notation $\langle\cdot\rangle$ stands for averaging over a
density matrix of the total system. We calculate averages over
nonequilibrium states where subsystems have a temperature gradient
or electron system is subjected to a finite voltage. To simplify
expressions, the following shorthand notations are introduced:
$\hat{P}=\int\hat{\Psi}^{\dagger}(r)\hat{p}_z\hat{\Psi}(r)dr$,
$\hat{Z}=\int\hat{\Psi}^{\dagger}(r)z\hat{\Psi}(r)dr$ and
$\hat{I}=\frac{e}{mL_z}\hat{P}$. The quantities are related by
$\dot{\hat{Z}}=\hat{P}/m$. The current (\ref{cur}) can be written
as
\begin{align}\label{cur2}
J_Q(t)
=&-\frac{2g}{m} \mathrm{Re}\sum_jT_jG_j^{<}(t,t),\\
G_j^{<}(t,t')\equiv&\langle\hat{P}(t)\hat{a}_j(t')\rangle.\nonumber
\end{align}
The energy transport problem is reduced to finding the "lesser"
Green's function $G_j^{<}(t,t')$. We will concentrate on a
steady-state situation where $G_j^{<}(t,t')=G_j^{<}(t-t')$.

The contour-ordered Green's function $G_j(\tau,\tau')$ can be
derived by the equation-of-motion technique.\cite{haug} 
Let us first consider the time-ordered Green's function
$G_j^{t}(t-t')$ at $T=0$. By differentiation and applying
Heisenberg's equation of motion we obtain
\begin{align}\label{eqmo2}
(i\partial_{t'}-\omega_j)G_j^t(t-t')=\frac{g}{\hbar}T_j\langle\hat{P}(t)\hat{Z}(t')\rangle^{t}.
\end{align}
The expression in brackets on the left hand side of
Eq.~(\ref{eqmo2}) can be interpreted as an inverse Green's function
$D_j^{t\,-1}$ of a free photon field. Thus Eq.~(\ref{eqmo2}) can be
solved by
 integration, yielding
\begin{align*}
G_j^t(t-t')=\frac{g}{\hbar}T_j\int
dt_1\langle\hat{P}(t)\hat{Z}(t_1)\rangle^{t} D_j^t(t_1-t').
\end{align*}
Using analytical continuation rules known as the Langreth's
theorem,\cite{haug} we obtain
\begin{align*}
G_j^<(t-t')=\frac{g}{\hbar}T_j&\int
dt_1\left[\langle\hat{P}(t)\hat{Z}(t_1)\rangle^{r}D_j^<(t_1-t')+\right.\nonumber\\
&\left.+\langle\hat{P}(t)\hat{Z}(t_1)\rangle^{<}D_j^a(t_1-t')\right].
\end{align*}
Superscripts $^r$, $^a$ and $^<$ stand for "retarded", "advanced"
and "lesser". For later purposes it is convenient to define the
Fourier transform
\begin{align*} \label{fgre}
&G_j^<(\omega)=\frac{1}{\hbar}g\,T_j\left[\langle\hat{P}\hat{Z}\rangle^{r}(\omega)
D_j^<(\omega)+ \langle\hat{P}\hat{Z}\rangle^{<}(\omega) D_j^a(\omega)\right]=\nonumber\\
&-\frac{gL_z^2\,m}{\hbar
e^2}T_j\left[\frac{i}{\omega}\langle\hat{I}\hat{I}\rangle^{r}(\omega)
D_j^<(\omega)+
\frac{i}{\omega}\langle\hat{I}\hat{I}\rangle^{<}(\omega)
D_j^a(\omega)\right].
\end{align*}
Now Eq.~\eqref{cur2} yields for the steady-state current
\begin{align}
J_Q=&\frac{2}{\hbar}\mathrm{Re}\sum_jT_j^2\int_{-\infty}^{\infty}
\frac{d\omega}{2\pi}\times\nonumber\\\times&\left[\frac{i\langle\hat{I}\hat{I}\rangle^{r}(\omega)}{\omega}
D_j^<(\omega)+
\frac{i\langle\hat{I}\hat{I}\rangle^{<}(\omega)}{\omega}
D_j^a(\omega)\right].
\end{align}
For a transmission line much longer than $\hbar \pi v/(k_B T)$, the
sum over the field modes can be replaced by integration according to
$\sum_j=\frac{L}{\pi}\int_0^{\infty} dk=\frac{L}{\pi
v}\int_{0}^{\infty} dw$. The current takes the form
\begin{align}\label{cur4}
J_Q=&\frac{2Z_0}{\pi }\mathrm{Re}\int  d\omega_j \hbar\omega_j
\int\frac{d\omega}{2\pi}\times\nonumber\\&\times\left[\frac{i\langle\hat{I}\hat{I}\rangle^{r}(\omega)}{\hbar\omega}
D_j^<(\omega)+
\frac{i\langle\hat{I}\hat{I}\rangle^{<}(\omega)}{\hbar\omega}
D_j^a(\omega)\right],
\end{align}
where $Z_0=\sqrt{l/c}$ is the characteristic impedance of the
transmission line.

The photon Green's functions at a finite temperature can be written
as $D_j^<=-2\pi i n_{\gamma}(\omega)\delta(\omega-\omega_j)$ and
$D_{j}^a=\frac{1}{\omega-\omega_j-i\eta}=\pi
i\delta(\omega-\omega_j)+P\frac{1}{\omega-\omega_j}$, where
$n_{\gamma}(\omega)$ is the Bose distribution. Inserting these
expressions into Eq.~(\ref{cur4}) gives
\begin{align}\label{cur5}
\!J_Q=2Z_0\int_0^\infty\frac{d\omega}{2\pi}\left[2\mathrm{Re}\langle\hat{I}\hat{I}\rangle^{r}(\omega)
n_{\gamma}(\omega)- \langle\hat{I}\hat{I}\rangle^{<}(\omega)
\right].
\end{align}
The correlators on the right hand side of Eq.~(\ref{cur5}) can be
expressed in terms of the noise power $S_I=\int_{-\infty}^{\infty}
e^{i\omega t}\langle \hat{I}(t)\hat{I}(0)\rangle dt$ as
$\langle\hat{I}\hat{I}\rangle^{<}(\omega)=S_I(-\omega)$ and
$\mathrm{Re}\langle\hat{I}\hat{I}\rangle^{r}(\omega)=\frac{1}{2}(S_I(\omega)-S_I(-\omega))$.
Expression (\ref{cur5}) is an exact formula for the energy flow and
valid even when the electron system is out of equilibrium. However,
it contains the exact current-current correlation functions of the
metallic island in the presence of the field. In equilibrium, these
are related to conductance through the Fluctuation-Dissipation
theorem and the Kubo formula. From formal point of view, the exact
expression for noise power determines the energy exchange process
completely. In the weak-coupling limit (the lowest order in
electron-photon coupling), one can neglect the field
and use the bare island correlators. 
On physical grounds one expects that the maximum energy transport is
achieved when the coupling is strong, thus a more accurate treatment
of the electron-photon interaction is desirable.

\begin{figure}[h]
\centering
\includegraphics[width=0.55\columnwidth]{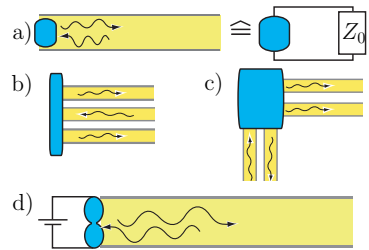}
\caption{(Color online) a) Circuit picture corresponding to the
studied system. Transmission line acts as a series resistor to the
island. In b) the parallel transmission lines couple to vertical
current noise and in c) the lines couple to vertical and horizontal
noise. The maximum heat conductance in b) is $G_Q$ and in c) is
$2G_Q$. The number of parallel transmission lines is irrelevant for
the maximum heat conductance. In d) the island contains a short
conductor and is externally biased.} \label{moni}
\end{figure}

The quadratic form of $H_{\gamma}$ and the linear coupling term
$H_{\mathrm{e-ph}}$, together with the density of states of a long
transmission line is precisely a Caldeira-Leggett representation of
an ohmic loss.\cite{legget} Solution to the equation of motion of
the voltage operator is
$\hat{V}_{\mathrm{TL}}(t)=\hat{V}_{\mathrm{TL}}^0(t)+Z_0\hat{I}(t)$,
where $\hat{V}_{\mathrm{TL}}^0(t)$ is the solution in the absence of
the island and $\hat{I}(t)$ is the current flowing in the electron
system. This notion microscopically motivates the circuit
approximation, where the transmission line can be thought as a
resistor in series with the metallic island, see Fig.~\ref{moni} a).

In the circuit description correlation functions can be calculated
by the Langevin approach, which allows us to relate the current
fluctuations in the presence of the environment to bare quantities.\cite{blanter} 
This yields
\begin{align}\label{cir}
\langle\hat{I}\hat{I}\rangle(\omega)=\frac{\langle\hat{I}\hat{I}\rangle_e(\omega)}{|1+G_e(\omega)Z_0|^2},
\end{align}
where $\langle\hat{I}\hat{I}\rangle_e(\omega)$ and $G_e(\omega)$ are
the current-current correlation function and conductance of the
island in the absence of the electromagnetic field. According to the
Kubo formula for conductance, the real part of the retarded function
appearing in Eq.~(\ref{cur5}) is related to the conductance as
$\mathrm{Re}
[G_e(\omega)]=\mathrm{Re}[\langle\hat{I}\hat{I}\rangle_e^{r}(\omega)]/(\hbar\omega)$.
With approximation \eqref{cir}, we then get
\begin{align}\label{master}
J_Q=&
\int_0^\infty\frac{d\omega}{2\pi}\frac{2Z_0}{|1+G_e(\omega)Z_0|^2}\times\nonumber\\
&\times\left\{[S_I^e(\omega)-S_I^e(-\omega)]n_{\gamma}(\omega)-
S_I^e(-\omega) \right\},
\end{align}
where $S_I^e(\omega)$ is the noise power for the isolated electron
system.

In (quasi)equilibrium the correlation functions are related through
a variant of Fluctuation-Dissipation formula
$S_I^e(\omega)=2\mathrm{Re}[G_e(\omega)]\hbar \omega n_e(\omega)$,
where $n_{\mathrm{e}}(\omega)$ is the Bose distribution function at
the electron temperature. Thus for this case 
\begin{align}\label{equi}
J_Q= \frac{4Z_0R_{\mathrm{e}}}{(R_{\mathrm{e}}+Z_0)^2 }
\int_0^\infty\frac{d\omega\hbar\omega}{2\pi}\left[
n_{\gamma}(\omega)-n_{\mathrm{e}}(\omega) \right],
\end{align}
when the island is assumed resistive $R_{\mathrm{e}}\equiv 1/G_e$.
Result (\ref{equi}) agrees with the one stated in Ref.~\onlinecite{schmidt} for heat flow between two resistors. 
After integration Eq.~(\ref{equi}) gives
\begin{align}
J_Q=r\frac{\pi^2k_B^2}{6h}(T_{\gamma}^2-T_{\mathrm{e}}^2),\quad r
\equiv \frac{4Z_0R_{\mathrm{e}}}{(R_{\mathrm{e}}+Z_0)^2 }
\end{align}
At small temperature difference this is just
\begin{equation}\label{tulos}
J_Q=rG_Q\Delta T,
\end{equation}
where $G_Q=\pi^2k_B^2T/3h$ is the universal quantum of heat
conductance \cite{pendry}. Thus, when $R_e=Z_0$ and thus $r=1$, 
the maximum one-channel heat transfer is
achieved. In the weak-coupling limit, where the exact correlation
functions in (\ref{cur5}) are replaced by bare correlators
$\langle\hat{I}\hat{I}\rangle_0(\omega)$, one recovers (\ref{tulos})
with the prefactor $Z_0R_{\mathrm{e}}/(R_{\mathrm{e}}+Z_0)^2$
replaced by $Z_0/R_{\mathrm{e}}$. Physically the weak coupling
result follows from the impedance mismatch $Z_0\ll R_{\mathrm{e}}$.

The above discussion can be generalized to incorporate several
photon channels realized by coupling the electron system to, say,
$N$ transmission lines, as in Fig.~\ref{moni}. Suppose that each
transmission line is described by a Hamiltonian of the form
(\ref{fotoni}) with the coupling (\ref{coupl}) corresponding to the
situation in Fig.~\ref{moni} b). The theoretical maximum heat
conductance for $N$ independent channels is $N\times G_Q$, but it is
not achieved in this case. An added transmission line does not
simply add an independent photon channel because it also effectively
acts as a series resistor in the coupling direction. Thus it
suppresses current fluctuations and affects the emitted energy in
all channels. The heat flow (\ref{tulos}) for multiple channels is
\begin{equation}\label{manyflow}
J_Q=\sum_i G_{Q}^i\frac{4R_{i}Z_i}{(R_{i}+\sum_i Z_i)^2 }\Delta T_i,
\end{equation}
where $Z_i$ is the characteristic impedance, $\Delta T_i$ the
temperature difference and $R_i$ the electron resistance associated
with line $i$. When all the transmission line fields are at the same
temperature, the maximum heat conductance given by
Eq.~(\ref{manyflow}) is still $G_Q$. Thus, adding parallel lines
does not increase this maximum. However, coupling the island to
perpendicular transmission lines, as in Fig.~\ref{moni} c), opens up
an independent transport channel. The difference in b) and c) is
that the lines in perpendicular directions couple to different
current components. The flow $J_Q$ is then a sum of two terms of the
form (\ref{manyflow}) 
and the maximum heat conductance is $2G_Q$. Similarly, coupling to
the remaining orthogonal direction yields the maximum heat
conductance 3$G_Q$.

\begin{figure}[h]
\centering
\includegraphics[width=0.8\columnwidth]{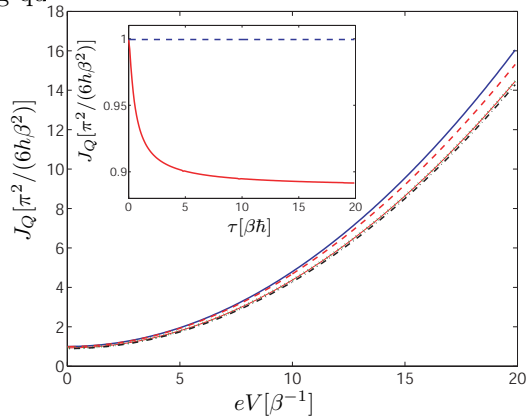}
\caption{(Color online) Energy flow from a symmetric chaotic cavity
$(N_L=N_R=1)$ as a function of voltage. We assumed that the cavity
conductance at zero frequency $G_e(0)$ is matched to $Z_0^{-1}$. The
different curves correspond to different cavity charge relaxation
times $\tau$, $\tau=0$ (solid), $\tau=0.1\hbar\beta$ (dashed),
$\tau=\hbar\beta$ (dotted), $\tau=10\hbar\beta$ (dash-dotted). Inset
shows the $\tau/\hbar\beta$-dependence of the heat flow from the
cavity ($V=0$), the horizontal dashed line corresponds to the heat
flow from an ideally matched ohmic resistor without frequency
dependence. When $\tau/\hbar\beta\ll1$ the heat flow is close to the
theoretical maximum and settles to a lower value as the fraction
increases.
 }
\label{jannite}
\end{figure}

Next we consider a case where the island contains a short contact
which is externally biased by potential difference, see Fig.
\ref{moni} d). Supposing that the electron transport is coherent and
neglecting interaction effects, current noise contains both the
equilibrium and shot noise and can be written as \cite{aguado}
\begin{equation}\label{noise}
\begin{split}
S_I(\omega)&=G_0\sum_m
T_m(1-T_m)\bigg[\frac{eV+\hbar\omega}{1-e^{\beta(-\hbar\omega-eV)}}\\&+
\frac{-eV+\hbar\omega}{1-e^{\beta(-\hbar\omega+eV)}} \bigg]
+G_0\sum_mT_m^2\frac{2\hbar\omega}{1-e^{-\beta\hbar\omega}},
\end{split}
\end{equation}
where $G_0=e^2/h$, $V$ is the bias voltage and $T_m$ is the
transmission eigenvalue of channel $m$. The sums of transmission
eigenvalues extend over the channel index and the spin. Inserting
expression (\ref{noise}) to the general formula (\ref{cur5}) we
discover
\begin{align}\label{nonequi}
J_Q^{\gamma}=r\left[\frac{1}{2}G_0\mathcal{L}_0(T_{\gamma}^2-T_{\mathrm{e}}^2)-\frac{1}{2}F_2GV^2\right],
\end{align}
where  $G=G_0\sum_mT_m=1/R_e$ is the island conductance,
$F_2=\sum_mT_m(1-T_m)/\sum_mT_m$ the Fano factor, and 
$\mathcal{L}_0=\pi^2k_{\mathrm{B}}^2/3e^2$ the Lorenz number. The
last term in $J_Q^{\gamma}$ corresponds to the increased emission by
shot noise.
The frequency dependence in Eq.~(\ref{noise}) is solely due to the
Fermi distribution and the emitted energy due to shot noise shows
only dependence on the bias voltage. Expression (\ref{noise}) is
valid only for low frequencies; generally $S_I(\omega)$ probes the
intrinsic (inverse) time scales of the conductor such as the time of
flight and the charge relaxation time. This is shown in
Fig.~\ref{jannite} where we have used the noise and conductance of
an interacting chaotic cavity \cite{hekking} to numerically compute
the energy flow.

Assume that the island is biased using superconducting wires with
contact conductances much higher than $G$. Such a setup provides
thermal insulation of the island \cite{meschke} while the voltage
still drops across the contact. The final temperature $T_e$ of the
island can be obtained from a heat balance equation where the Joule
heating $J_Q^J$ from the voltage source is balanced by heat flow to
photons and phonons as
$J_Q^J+J_Q^{\gamma}+J_Q^{\mathrm{ph}}=GV^2+J_Q^{\gamma}+\Sigma\Omega[T_\mathrm{ph}^5-T_e^5]=0,$
where $\Sigma$ is the electron-phonon coupling constant
\cite{giazotto} and $\Omega$ is the volume of the island.
There is a crossover temperature $T_{\rm cr}=(rG_0
\mathcal{L}_0/(2\Sigma \Omega))^{1/3}$ below which the photon
transport is the dominant process. For example, with the parameters
of Ref.~\onlinecite{meschke}, $T_{\rm cr}$ would be roughly 140 mK
-- for smaller objects such as carbon nanotubes, it could be made
larger at least by one order of magnitude. Much below $T_{\rm cr}$
the final electron temperature is
\begin{equation}
T_e=\sqrt{ T_{\gamma}^2+\frac{T_{\rm ph}^5}{T_{\rm
cr}^3}+\frac{2G}{G_0r}\left(1-\frac{F_2r}{2}\right)\frac{V^2}{\mathcal{L}_0}}
\end{equation}
and above the crossover it is
\begin{equation}
T_e=\left( T_{\mathrm{ph}}^5+T_{\rm cr}^3
T_\gamma^2+G\left(1-\frac{F_2r}{2}\right)\frac{V^2}{\Sigma
\Omega}\right)^{\frac{1}{5}}.
\end{equation}
In both limits the Joule heating is reduced by the factor
$(1-F_2r/2)$ because a fraction of it flows to photons. In the case
of an ideally matched ($r=1$) tunnel junction ($F_2=1$), exactly
half of the Joule heat goes to photons. In experiments the
parameters $V$, $T_{\rm ph}$, $T_\gamma$ and $r$ can be varied in
situ to investigate the photon transport contribution, as
demonstrated in Ref.~\onlinecite{meschke} for $r$, $T_{\rm ph}$ and
$T_\gamma$.

In conclusion, we studied a microscopic model of photon transport in
nanostructures using a Green's-function method and derived a general
expression for the energy flow between a metallic island and a
transmission line field. We showed how electron and photon transport
are related through frequency-dependent current noise. We
demonstrated efficiency of the energy flow formula by deriving
quantized photon heat conductance and studying effects of electron
shot noise to photon transport. We propose to measure the shot-noise
effect illustrated as the voltage-dependent term in
Eq.~\eqref{nonequi} by modifying the setup in
Ref.~\onlinecite{meschke} to include a small mesoscopic junction.

We thank Jukka Pekola, Matthias Meschke and Henning Schomerus for
insightful discussions. TTH acknowledges the Academy of Finland for
funding.

\end{document}